\documentclass[preprint,prb,superscriptaddress,byrevtex,showpacs, 
showkeys,11pt,onecolumn,showkeys,nobalancelastpage]{revtex4}
\usepackage{amsmath}
\usepackage{amsfonts}

\input{tcilatex}
\begin{document}

\preprint{}
\title[Short title for running header]{PbCu$_{3}$TeO$_{7}$: An $S=\frac{1}{2}
$\ staircase Kagome lattice with significant intra- and inter-plane couplings%
}
\author{B. Koteswararao}
\affiliation{Center of Condensed Matter Sciences, National Taiwan University, Taipei
10617, Taiwan}
\author{R. Kumar}
\affiliation{Department of Physics, Indian Institute of Technology Bombay, Mumbai 400076,
India}
\author{ Jayita Chakraborty}
\affiliation{Department of Solid State Physics, Indian Association for the Cultivation of
Science, Jadavpur, Kolkata 700 032, India}
\author{Byung-Gu Jeon}
\affiliation{CeNSCMR, Department of Physics and Astronomy, Seoul National University,
Seoul 151-747, Republic of Korea}
\author{A. V. Mahajan}
\affiliation{Department of Physics, Indian Institute of Technology Bombay, Mumbai 400076,
India}
\author{I. Dasgupta}
\affiliation{Department of Solid State Physics, Indian Association for the Cultivation of
Science, Jadavpur, Kolkata 700 032, India}
\author{Kee Hoon Kim}
\affiliation{CeNSCMR, Department of Physics and Astronomy, Seoul National University,
Seoul 151-747, Republic of Korea}
\author{F. C. Chou}
\email{fcchou@ntu.edu.tw}
\affiliation{Center of Condensed Matter Sciences, National Taiwan University, Taipei
10617, Taiwan}
\affiliation{National Synchrotron Radiation Research Center, Hsinchu 30076, Taiwan}
\keywords{PbCu$_{3}$TeO$_{7}$, frustration, Kagome, quantum spin}
\pacs{71.27.+a 75.10.Jm 75.50.Ee }

\begin{abstract}
We have synthesized polycrystalline and single crystal samples of PbCu$_{3}$%
TeO$_{7}$ and studied its properties via magnetic susceptibility $\chi (T)$
and heat-capacity $C_{p}(T)$ measurements and also electronic structure
calculations. Whereas the crystal structure is suggestive of the presence of
a quasi-$2D$ network of Cu$^{2+}$ $(S=1/2)$ buckled staircase Kagome layers,
the $\chi (T)$ data show magnetic anisotropy and three magnetic anomalies at
temperatures, $T_{N1}\sim 36$ K, $T_{N2}\sim 25$ K, $T_{N3}\sim 17$ K,
respectively. The $\chi (T)$ data follow the \ Curie-Weiss law above $200$ K
and a Curie-Weiss temperature $\theta _{CW}\sim -150$ K is obtained. The
data deviate from the simple Curie-Weiss law below $200$ K, which is well
above $T_{N1}$, suggesting the presence of competing magnetic interactions.
The magnetic anomaly at $T_{N3}$ appears to be of first-order from
magnetization measurements, although our heat-capacity $C_{p}(T)$ results do
not display any anomaly at $T_{N3}.$ The hopping integrals obtained from our
electronic structure calculations suggest the presence of significant
intra-Kagome (next-nearest neighbor and diagonal) and inter-Kagome
couplings. These couplings take the PbCu$_{3}$TeO$_{7}$ system away from a
disordered ground state and lead to long-range order, in contrast to what
might be expected for an ideal (isotropic) $2D$ Kagome system.
\end{abstract}

\volumeyear{year}
\volumenumber{number}
\issuenumber{number}
\eid{identifier}
\date[Date text]{date}
\received[Received text]{date}
\revised[Revised text]{date}
\accepted[Accepted text]{date}
\published[Published text]{date}
\startpage{1}
\endpage{2}
\maketitle

\section{\textbf{Introduction}}

Frustrated magnetism in Kagome Heisenberg antiferromagnetic systems (KHAF)
has been a major subject in condensed matter physics due to their
unconventional, exotic ground states which emerge from the interplay between
geometrical frustration and low-dimensional quantum effects.\cite%
{GFM2005,GFM2011} In particular, quantum fluctuations for $S=1/2$ systems
are found to be strong among KHAF and can lead to interesting behavior like
that of a spin liquid.\cite{L. Balents Nature 2010} Theoretical studies on
ideal $S=1/2$ isotropic KHAF lattice have demonstrated that it has a
Resonating Valence Bond (RVB)-like disordered ground state.\cite{S
SachdevPRB1992,J T Chalker1992} Recent numerical studies \cite{SYan2011}
have also predicted that its ground state is a spin liquid with a small
spin-gap ($\Delta /k_{B}$) of $(0.03\sim 0.05)J/k_{B}$ to its triplet
excited state (where $J$ is the exchange interaction between
nearest-neighbor ($nn$) spins). A limited number of experimental
realizations of structurally ideal, $S=1/2,$ KHAF have been found, which
include Zn and Mg-Herberthsmithite, Kapellasite, and Haydeeite.\cite%
{lists=1/2kagome,E. Kermarrec2011,B. Fak 2012} Among these,
Zn-Herberthsmithite ZnCu$_{3}$(OH)$_{6}$Cl$_{2}$ is the best example to
explain isotropic $S=1/2$ KHAF behavior. Experimental studies on
Zn-Herberthsmithite have shown that there is no magnetic ordering down to $%
\frac{J/k_{B}}{3000}$ K, which implies that it has a disordered ground state.%
\cite{PMendelsPRL2007} But an unambiguous proof for the existence of a
spin-gap from an experimental point of view is lacking. The natural ion
exchange of Cu and Zn or Mg is inevitable in these systems, which is
probably the main obstacle to detect the intrinsic nature of these $S=1/2$
KHAF. On the other hand, there are a few anisotropic KHAF systems such as
Volborthite Cu$_{3}$V$_{2}$O$_{7}$(OH)$_{2}$.2H$_{2}$O \cite{M. A.
Lafontaine JSSC 1990} and Vesignieite BaCu$_{3}$V$_{2}$O$_{8}$(OH)$_{2}$ 
\cite{YOkamoto 2009}, which do not have much atomic site-disorder, in which
the Kagome layers are built by two nearest neighbor ($nn$) exchange
couplings. These have also been studied to understand the real ground state
physics of a Kagome system. Despite the presence of significant anisotropy,
these systems show interesting spin dynamics.\cite{R. H. ColmanPRB2011}
There is another kind of anisotropic Kagome lattice in Rb$_{2}$Cu$_{3}$SnF$%
_{12},$\cite{K. Morita JPSJ 2008} where the Kagome layer is formed by four
exchange couplings and has a spin-gap of $20$ K. The pinwheel valence bond
solid (VBS) excitations are realized in this anisotropic KHAF system.\cite%
{K. Matan Nature Phys2010}\textbf{\ }Theoretical predictions also suggest
that when the Kagome layer is perturbed with additional exchange couplings
such as next-nearest neighbor $nnn$ and diagonal couplings, the system is
driven to a novel ordered state from the disordered state of isotropic
Kagome layer.\cite{A. B. Harris PRB1992} These predictions\textbf{\ }%
prompted us to look for newer $S=1/2,$ KHAF systems to explore interesting
physics associated with the presence of anisotropy and additional couplings.

Herein, we introduce a new anisotropic magnetic system, PbCu$_{3}$TeO$_{7},$ 
\cite{B WedelZNB1996} which has $S=1/2$ staircase (buckled) Kagome layers
built by Cu and O atoms (see Fig. $1$). This system has Kagome layers
similar to those of A$_{3}$V$_{2}$O$_{8}$ (A=$\beta $-Cu, Ni, Co). \cite%
{Cu3V2O8,Ni3Co3synthesis} \ Detailed studies have been done on single
crystals of $S=1$ and $S=3/2$ staircase Kagome systems Ni$_{3}$V$_{2}$O$_{8}$
and Co$_{3}$V$_{2}$O$_{8}.$ The Ni$_{3}$V$_{2}$O$_{8}$ system has four
competing magnetic transitions (two incommensurate and two commensurate)
below $9$ K, one of which is a multiferroic transition at $T\sim 6$ K. \cite%
{GLawesPRL2004,G LawesPRL2005} However, Co$_{3}$V$_{2}$O$_{8}$ does not show
multiferroicity, which might be due to its high spin state with low quantum
fluctuations. Less work has been done on the $S=1/2$ analog $\beta $-Cu$_{3}$%
V$_{2}$O$_{8},$ probably due to non availability of single crystals.

We report here the magnetic properties of PbCu$_{3}$TeO$_{7}$. The magnetic
susceptibility $\chi (T)$ data indicate that the dominant exchange
interactions between Cu$^{2+}$ ions are antiferromagnetic (AF) with a
Curie-Weiss temperature ($\theta _{CW}$) of about $-150$ K. The $\chi (T)$
deviates from Curie-Weiss behavior below $\sim 200$ K. We observed a total
of three magnetic anomalies ($T_{N1}\sim 36$ K, $T_{N2}\sim 25$ K, and $%
T_{N3}\sim 16$ K) in the $\chi (T)$ data of a single crystal. The magnetic
anomalies at $T_{N1}$ and $T_{N3}$ were evident only when the applied
magnetic field $H$ was parallel to the crystallographic $a$-axis, whereas
the one at $T_{N2}$ could be observed only for $H\parallel c$. The anomaly
at $T_{N3}$ is first order in nature and is field sensitive. On the other
hand, heat capacity data in zero field (for both polycrystalline and single
crystal samples) showed anomalies of $T_{N1}$ and $T_{N2}$ only. The
first-order transition at $T_{N3}$ could not be observed by us in heat
capacity data. We suggest that this might be due to the small latent heat
involved in this transition. Our\ electronic structure calculations in
conjunction with our experimental findings suggest the presence of various
competing magnetic couplings, in addition to $nn$, in PbCu$_{3}$TeO$_{7}$
which in turn cause a deviation from the superficially anticipated regular
Kagome geometry leading to long-range order (LRO).

\section{\textbf{Experimental details}}

Polycrystalline samples of PbCu$_{3}$TeO$_{7}$ were prepared by conventional
solid-state reaction method using PbO, CuO, and Te precursors. The
stoichiometric amount of chemicals were ground thoroughly and fired at 750 $%
^{o}$C for $5$ days with three intermediate grindings. To obtain single
crystals, a mixture of PbCu$_{3}$TeO$_{7}$ and NaCl/KCl flux in the mass
ratio $1:2$ was charged in an alumina crucible, fired at $800$ $^{o}$C for $%
24$ hrs and then cooled slowly to $650$ $^{o}$C with a cooling rate of $1$ $%
^{o}$C per hour. Single crystals were extracted after washing the flux with
hot water. The\ x-ray diffraction (XRD) data were collected using $D8$
Advance (Bruker) on the single crystal and on (polycrystalline) powder
obtained from crushing the crystals. Magnetic and heat-capacity measurements
were performed on polycrystalline and single crystal samples using a
Physical Property Measurement System (PPMS) from Quantum Design. For
heat-capacity, we employed the thermal relaxation method with two time
constants ($2\tau $ model) to fit the heat-capacity data. The single crystal
used for the magnetization measurements is of mass $2$ mg and approximate
dimensions $1.5$ mm x $0.5$ mm x $0.45$ mm as shown in Fig. $2(b)$.

\section{\textbf{Results and Discussion}}

\subsection{X-ray diffraction and structural features}

The single phase nature of the polycrystalline sample was confirmed by
comparing the measured XRD pattern with the calculated one generated by
powdercel\cite{powdercel} using the initial structural parameters of the
orthorhombic space group \textit{Pnma} (space group No. $62$) given by B.
Wedel, \textit{et al.}\cite{B WedelZNB1996} Rietveld refinement of the XRD
pattern was done using the Fullprof suite software\cite{Carvajal 1993} (as
shown in Fig. $2(a)$). The lattice parameters obtained from the refinement
are $a=10.484$ \AA , $b=6.347$ \AA , and $c=8.807$ \AA .\ The obtained
atomic positions and occupancies are summarized in Table $1$. These
structural parameters are in good agreement with the previously published
values.\cite{B WedelZNB1996} A rectangular crystal was placed on a glass
slide with the top face (see Fig. $2$) parallel to the plane of the slide
and XRD was performed. It resulted in the appearance of only $(0ll)$ peaks
as shown in Fig. $2(b)$). This indicates that the top crystal surface\ (as
indicated in Fig. 2) is perpendicular to the $[0ll]$ direction or $bc-$%
plane. This means that the $a-$axis lies in the as-grown plane of the
crystal. Note that crystallographic $a-$axis is perpendicular to the
staircase Kagome layers, as shown in Fig. $1(b)$. We have further taken the
Laue pattern of the crystal, and found that the crystallographic $a-$axis is
along the length of the crystal. The other crystallographic axes ($b$ and $c 
$) were also identified and make roughly $45^{o}$ with the top plane of the
crystal, as shown in the inset of Fig. $2(b)$.

The staircase Kagome planes in the PbCu$_{3}$TeO$_{7}$\cite{B WedelZNB1996}
structure are formed by Cu atoms (Cu1 and Cu2) as shown in Fig. $1$. The Cu$%
^{2+}$ ions are coupled via O$^{2-}$ ions and form buckled layers in the $bc$%
-planes, as shown in Fig. $1(b)$. According to the Cu-Cu bond-lengths and
the Cu-O-Cu bond-angles, the staircase Kagome plane possibly has four
different nearest neighbor $(nn)$ exchange constants $J_{1}$, $J_{2}$, $%
J_{3} $ and $J_{4}$ (see Fig. $1(b)$). In this staircase Kagome plane, Cu1
atoms form linear chains with the couplings alternating between $J_{2}$ and $%
J_{4}$. These Cu1 chains are connected with each other via Cu2 atoms with
couplings $J_{1}$ and $J_{3}$ to form a $2D$ network. Other possible
magnetic coupling paths are also shown in Fig. $1(f)$, the hopping integrals
for which have been calculated by us. These are $t_{5},$ $t_{6},$ $t_{7},$
and $t_{8}$ which are in addition to the $nn$ hoppings $t_{1},$ $t_{2},$ $%
t_{3}$ and $t_{4}$ (corresponding to $J_{1}$, $J_{2}$, $J_{3}$ and $J_{4}$,
respectively). The Kagome planes are separated by Pb and Te atoms with a
small inter-planar distance of about $4$ \AA , which suggests that
significant three-dimensional couplings between the layers might be present.

\subsection{\textbf{Magnetic measurements:}}

The magnetization $M$ was measured as a function of temperature $T$. The
dependencies of the magnetic susceptibility $\chi (=M/H)$ in the range from $%
2$ K to $370$ K for the polycrystalline and single crystal samples of PbCu$%
_{3}$TeO$_{7}$ are shown in Fig. $3(a)$. Magnetic measurements were
performed on the $2$ mg single crystal (shown in Fig. $2(b)$) for each of
the crystal direction $a$, $b,$ and $c,$ respectively in an applied magnetic
field $H$. The $\chi (T)$ data follow a Curie-Weiss behavior in high-$T$
region. Even in the paramagnetic region, a significant anisotropy is seen
between the field orientation along $a$ and the other two perpendicular
directions ($b$ and $c)$. This means that the intra-Kagome ($bc-$plane)
anisotropy (difference between $b-$ and $c-$ axis) is small compared to the
inter-Kagome anisotropy. The $\chi (T)$ data of\ the polycrystalline sample
lie in between the single crystal data for $H\parallel a$ and $H\parallel
bc, $ but for $T>200$ K they are close to the single crystal data of $%
H\parallel a,$\ while below $200$\ K they are closer to the single crystal
data for $H\perp a$. The temperature independent magnetic susceptibility $%
\chi _{0}$ was estimated from polycrystalline data at high$-T$ from $300$ K
to $800$ K (data not shown here). We subtracted $\chi _{0}=-(3\pm 0.3)\times
10^{-5}$ cm$^{3}$/mol Cu from our measured data and then plotted $(\chi
-\chi _{0})^{-1} $\ vs $T$ in Fig. $3(b)$. The core diamagnetic
susceptibility ($\chi _{core}$) of PbCu$_{3}$TeO$_{7}$ is calculated to be $%
-5.23\times 10^{-5}$ cm$^{3}$/mol Cu from $\chi _{core}$ of individual ions.%
\cite{P. W. Selwood 1956} The Van Vleck paramagnetic susceptibility is then
estimated to be $\chi _{vv}=\chi _{0}-\chi _{core}\sim (2.23\pm 0.3)\times
10^{-5}$ cm$^{3}$/mol Cu, which is comparable with other cuprates.\cite{N
MotoyamaPRL1996} The $\chi (T)$ data follow the Curie-Weiss (CW) law $\left( 
\frac{C}{T-\theta _{CW}}\right) $ in the $T-$range $200$ - $370$ K where $C$
is the Curie constant and $\theta _{CW}$\ is the Curie-Weiss temperature.
The obtained parameters from the single crystal data are given in the Table
II. The $C$ value corresponds to an effective magnetic moment ($\mu _{eff})$
of about $2$ $\mu _{B}$/Cu, which is slightly larger than the spin-only
value for $S=1/2$ which is $1.73$ $\mu _{B}$. The negative $\theta _{CW}$
indicates that the dominant couplings are antiferromagnetic in nature. There
is a deviation from the Curie-Weiss law below $200$ K, which indicates that
it is not a simple paramagnet at low$-T$. This deviation might be a result
of the competitive exchange couplings present in the quasi-$2D$ staircase
Kagome planes. The $\chi (T)$ data of polycrystalline samples exhibit a
total of three magnetic anomalies at $T_{N1}\sim 36$ K, $T_{N2}\sim 25$ K,
and $T_{N3}\sim 17$ K, respectively. The ordering temperature ($T_{N1}$) is
relatively smaller than $\theta _{CW}$ and the temperature below which a
deviation from CW behavior is seen ($200$ K), which indicates that the
system is moderately frustrated. These three magnetic anomalies ($T_{N1}$, $%
T_{N2}$ and $T_{N3}$) are also seen in the single crystal data, but with
significant anisotropy. All of the anomalies are not evident for each field
direction. For instance, the magnetic anomaly\ $T_{N1}$ is evident only when 
$H\parallel a$, and the anomaly at $T_{N2}$ is seen only for $H$ $\parallel
c $. On the other hand \ for $H$ $\parallel b$, no clear anomalies were
observed. The anomaly at $T_{N3}$ is prominent for $H\parallel a$ (see Fig. $%
4$). Although, the appearance of anomalies in the $\chi (T)$ data depends on
the field direction, the transitions are not field driven because the two
anomalies ($T_{N1}$ and $T_{N2}$) appear in zero-field heat capacity data as
will be discussed in a later section. Similar $H-$direction dependent
anomalies were also reported in a multiferroic, spiral magnet FeVO$_{4}$.%
\cite{A Daoud-Aladine2009} The susceptibility $\chi _{b}$ \ ($\chi (T)$ for $%
H$ $\parallel b$) increases below the transition while $\chi _{a}$ and $\chi
_{c}$ decrease below the transition. This suggests that $b-$axis might be
the hard-axis of magnetization. We have also performed dielectric constant
and electric polarization versus $T$ down to $2$ K on the polycrystalline
samples, neither anomaly in the dielectric constant nor electric
polarization are observed below the transition temperatures.

In order to obtain insight about the transition at $T_{N3}$, we have
measured $M$ of a\ collection of \ carefully oriented crystals ($H\parallel
a $), while cooling and warming in different $H^{\prime }s$ of $1$ kOe, $5$
kOe, $10$ kOe, $15$ kOe, respectively, as shown in Fig. $4$. The transition
at $T_{N3}$ is sharp and looks different from those of at $T_{N1}$ and $%
T_{N2}$. There is a difference in the position of the peak for the
zero-field-cooled warming (ZFCW) data and the field-cooled cooling (FCC)
data, while no difference is observed between ZFC warming and FC warming
(FCW) data. This kind of thermal hysteresis is a characteristic feature of \
a first-order phase transition.\cite{R NirmalaPRB2007} Similar sharp,
first-order anomalies are also observed at $3.9$ K in Ni$_{3}$V$_{2}$O$_{8}$%
\cite{GLawesPRL2004} and $6$ K in Co$_{3}$V$_{2}$O$_{8}$ \cite{Yukio YASUI
JPSJ 2007} staircase Kagome systems. Moreover, the observed peak position
moves to higher temperatures with increasing $H$ and this variation
(summarized in Table III) suggests that the observed first-order transition
is ferromagnetic in nature. The $T_{N3}$ peak also broadens and the
difference between the position (in temperature) of the warming and cooling
peak decreases as $H$ increases. This kind of field-induced broadening is
also a characteristic feature of the\ first-order transition.\cite{D P Rojas
PRB2011} Finally, the $T_{N3}$ peak is suppressed in a field of about $20$
kOe, however $T_{N1}$ and $T_{N2}$ peaks remain, more or less, unaffected by
magnetic fields upto $70$ kOe and do not exhibit any thermal hysteresis like
observed for $T_{N3}$ (data not shown here).

\subsection{\textbf{Heat-Capacity measurements}}

The $T-$dependent heat-capacity $C_{p}$ of the polycrystalline PbCu$_{3}$TeO$%
_{7}$ samples in the range from $2$ to $300$ K is presented in Fig. $5$. The 
$C_{p}(T)$ data were measured using the PPMS by the thermal\ relaxation
method. For a magnetic insulator, one expects both lattice and magnetic
contributions to the heat-capacity. We used the Debye model\cite{kittel} to
obtain the lattice part in the following manner. We fit the $C_{p}(T)$ data
in the $T-$range from $150$ K to $300$ K with a linear combination of two
Debye integrals as given below

\begin{equation}
C_{p}(T)=9rNk_{B}\dsum\limits_{i=1,2}C_{i}\left( \frac{T}{\theta _{D}^{i}}%
\right) ^{3}\dint\limits_{0}^{x_{D}^{i}}\frac{x^{4}e^{x}}{(e^{x}-1)^{2}}dx
\label{debye}
\end{equation}

Here $r$ is the number of atoms per formula unit, $\theta _{D}^{i}$ is a
Debye temperature, and $x_{D}^{i}=\theta _{D}^{i}/T$ ($i=1,2$). The\ fitting
yields $C_{1}=0.43\pm 0.05,$ $\theta _{D}^{1}=(280\pm 10)$ K$,$ $%
C_{2}=0.5\pm 0.05,$ and $\theta _{D}^{2}=772\pm 30$ K. The fitting curve was
extrapolated down to $2$ K and this was then subtracted from the measured $%
C_{p}(T).$ Magnetic heat-capacity $C_{m}(T)$ was thus obtained (see inset of
Fig. $5$). Sharp anomalies are observed at$\ 36$ K and $25$ K, which agree
well with the anomalies found at $T_{N1}$ and $T_{N2}$ in $\chi (T)$.\ We
observed the presence of $C_{m}(T)$ upto $150$ K which is well above $T_{N1}$%
, which signifies the presence of magnetic correlations above $T_{N1}$ and
this behavior is also consistent with the magnetic data. However, we could
not detect any transition at $T_{N3}$ in the data. In fact, detailed $%
C_{p}(T)$ measurements were done around $16$ K with a large number of
points. The raw data of the sample temperature as a function of time at each
of these temperatures showed a good fit to the $2\tau $ model. \ In the case
of a first-order transition, it has been documented that deviations from the
simple $2\tau $ behavior occur for $C_{p}$ data at the transition
temperature.\cite{J C Lashley CpPPMSsensitivity} In light of the missing
supporting evidence for $T_{N3}$ from the $C_{p}(T)$ measurement, one may
suspect that the existence of $T_{N3}$ from susceptibility data may not be
intrinsic. On the other hand, the inability to detect similar kind of
first-order phase transition has also been argued in the literature.\cite{J
C Lashley CpPPMSsensitivity} Some other technique like the adiabatic method
or a better pulse-sequence design\cite{R.W. NewsomeRSI2004} may be needed
for measuring small latent heat or entropy change transitions. Although our
heat capacity measurements do not support the existence of the $T_{N3}$
phase transition, we must stress that the $T_{N3}$ anomaly has been
consistently observed in the magnetization measurements for different
batches of powder samples as well as single crystals, i.e., the possibility
of contribution from external magnetic impurity phase due to different
preparation conditions or contamination can be ruled out. More sensitive
heat capacity measurement are being planned to clarify this issue. Overall
we conclude that the intrinsic origin of this transition is an open question.

The entropy change $S_{M}$ calculated from the magnetic heat-capacity is
about $5.63$ J/mol K Cu, which is in good agreement with the expected $R\ln
(2S+1)$\ ($5.76$ J/mol K) for $S=%
%TCIMACRO{\U{bd}}%
%BeginExpansion
{\frac12}%
%EndExpansion
$ systems. The $S_{m}$ value at transition $T_{N1}$ is found to be $2.88$
J/mol K Cu, which is $50\%$ of the total entropy and the rest of the entropy
is released in the paramagnetic region well above $T_{N1}$, suggesting the
presence of frustrated correlations. The observation of significant entropy
well above ordering temperature is generally observed in frustrated spin
systems.\cite{L. K. Alexander 2007} Additionally, nearly all the entropy
change has taken place by the time one approaches (with decreasing
temperature) the $T_{N3}$ transition. This is the reason I presume that $%
T_{N3}$ is a weak transition such as a slight canting from the AF ordered
state to produce weak FM moment. This might explain the difficulty in
observing the transition in heat capacity.

\subsection{Electronic structure calculations}

In order to study the electronic structure of PbCu$_{3}$TeO$_{7},$ we have
carried out first principles density functional theory (DFT) calculations
within the local-density approximation (LDA) by employing the Stuttgart
TBLMTO-47 code based on the linear muffin-tin orbital (LMTO) method in the
atomic sphere approximation (ASA).\cite{Anderson} The basis set for the
self-consistent electronic structure calculations for PbCu$_{3}$TeO$_{7}$ in
TB-LMTO ASA includes Pb (s, p), Cu (s, p, d), Te (s, p) and O (s, p) and the
rest are downfolded. A (4, 8, 4) $k$-mesh has been used for
self-consistency. All the $k$-space integrations were performed using the
tetrahedron method. In order to ascertain the accuracy of our ASA
calculations we also performed the electronic structure calculation using
projected augmented wave (PAW) \cite{blochl} method encoded in the Vienna $%
ab $-initio simulation package(VASP).\cite{vasp} The density of states
calculated by these two different approaches is found to agree well with
each other. In order to extract various hopping integrals between the Cu
atoms, we have employed the N$^{th}$ order muffin tin orbital (NMTO)
downfolding method.\cite{Saha-Dasgupta,Tank,Saha}

The non-spin-polarized band structure for PbCu$_{3}$TeO$_{7}$ is displayed
in Fig.~$6$. The bands are plotted along the various high symmetry points of
the Brillouin zone corresponding to the orthorhombic lattice. All the
energies are measured with respect to the Fermi level of the compound. The
characteristic feature of the non-spin-polarized band structure displayed in
Fig. $6$ is an isolated set of twelve bands crossing the Fermi level and
these bands are predominantly derived from the antibonding linear
combination of Cu-d$_{x^{2}-y^{2}}$ and O-p states. These twelve bands are
well separated from the low-lying O-p and other Cu -d valence bands. This
isolated Cu-d$_{x^{2}-y^{2}}$ twelve band complex is responsible for the
low-energy physics of this compound. Fig.~$7$ shows the non-spin-polarized
density of states (DOS) of PbCu$_{3}$TeO$_{7}$. As expected, there is strong
hybridization between the O-p and Cu-d states. The occupied Pb-6s states lie
below the Fermi level and empty Pb-p and Te-s states are above the Fermi
level.

We have employed the NMTO downfolding method to map our LDA results to a
low-energy orthogonal tightbinding Hamiltonian by integrating out the high
energy degrees of freedom from the all-orbital LDA calculation. The Fourier
transformation of the downfolded Hamiltonian $H=\sum_{ij}t_{ij}\left(
c_{i}^{\dagger }c_{j}+h.c.\right) $ gives the various hopping integrals $%
t_{ij}$ between the Cu atoms. These hopping integrals will determine the
dominant exchange paths for the system. For the present compound we have
kept only the Cu-d$_{x^{2}-y^{2}}$ orbital for each Cu atom in the unit cell
in the basis and integrated out all the rest to derive the low-energy model
Hamiltonian. We show the downfolded bands in Fig. $6$ in comparison to the
full LDA band structure and the agreement between the two is found to be
excellent. The hopping integrals ($>$ 10 meV) obtained from the NMTO
downfolding method are listed in Table IV and V and the exchange paths are
indicated in Fig. $1(f)$ and $1(g)$.

The strongest hopping in the Kagome plane is $t_{3}$ between the
corner-sharing Cu1 and Cu2 ions. This hopping is primarily mediated by O2
situated in the basal plane of the Cu1 octahedron forming a strong $pd\sigma 
$ antibond with the Cu1 $d_{x^{2}-y^{2}}$ orbital. The $nn$ hopping $t_{1}$
between the edge sharing Cu1 octahedron (see Fig. $1(c)$) and Cu2
tetrahedron (see Fig. $1(d)$), is found to be much smaller than $t_{3}$. In
order to obtain further insights, we have plotted the Wannier function of
Cu2 $d_{x^{2}-y^{2}}$ orbital corresponding to the $t_{3}$ and $t_{1}$
hoppings (see Fig. $8$). The plot reveals that the Cu2 $d_{x^{2}-y^{2}}$
orbital forms strong $pd\sigma $ antibonds with the neighboring oxygens. We
can see that the weight of Wannier function at Cu1 site which is at a
distance $3.264$ \AA ~ is large compared to the weight at Cu1 site which is
at a distance $2.90$~\AA ~indicating $t_3$ hopping will be stronger in
comparison to $t_1$ . The bond angle of Cu1-O2-Cu2 corresponding to the $%
t_{3}$ hopping is $118.06^{\circ }$, while the bond angle of Cu1-O1-Cu2 is $%
93^{\circ }$ for the $t_{1}$ hopping path. The strength of antiferromagnetic
interactions are strongly dependent on the angle between the bonds. When the
angle is close to $90^{\circ }$, the AF super-exchange process is suppressed
due to the orthogonality of Cu 3d and O 2p orbitals. So the exchange
coupling along the path $t_{1}$ is expected to be ferromagnetic. For $t_{4}$
hopping, two Cu1 octahedra are corner-shared with each other and the
bond-angle of Cu1-O3-Cu1 is $107^{\circ }$ while for the $t_{2}$ hopping
path, the two Cu1 octahedra share an edge with each other and the bond angle
of Cu1-O2-Cu1 is $105^{\circ }$. As argued earlier, the $t_{4}$ hopping
dominates over the $t_{2}$ hopping path due to the corner sharing topology
of the Cu1 octahedra. The second strongest hopping is $t_{7}$ between the
Cu2 atoms. This Cu2-Cu2 hopping primarily proceeds via the oxygens. As a
result, the strength of the Cu-O4-O4-Cu spin exchange is primarily governed
by the O-O distance and in this case O-O distance is 2.57 \AA ~(smaller than
the van der Waals distance). Moreover, the bond-angle of Cu-O4-O4 is about $%
167^{o},$ which makes this hopping stronger than some of the $nn$ hoppings ($%
t_{2},$ $t_{3},$ and $t_{4}$).

The hopping paths perpendicular to the Kagome plane are listed in Table V.
Since the distances between the Cu ions in two different Kagome plane is
small, it is expected that there is substantial amount of hoppings between
the Kagome planes. The strongest interaction between the Kagome plane is $%
t_{2}^{i}$. This interaction is mediated via Te following the path
Cu-O-Te-O-Cu.

Our electronic structure calculations reveal that intra-kagome and
inter-kagome exchange interactions are long ranged. In general, an ideal 2D
Kagome system with only uniform $nn$ couplings does not order even at zero$%
-T $. However, spin wave theory\cite{A. B. Harris PRB1992} predicts that the
presence of $2^{nd}$ $nn$ ($J_{nnn}$) and $3^{rd}$ $nn$ ($J_{diagonal}$)
couplings in the Kagome plane can drive the system to an ordered ground
state. When $J_{diagonal}>J_{nnn}$, $\sqrt{3}\times \sqrt{3}$\ N\'{e}el
ordered state is favored, while the $q=0$\ N\'{e}el state occurs in the case
of $J_{nnn}>J_{diagonal}.$\ A rich theoretical magnetic phase diagram has
also been built based on the type (antiferro and ferromagnetic) and the
relative strength ($J_{nnn}/J_{diagonal}$) of interactions in the $2D $
Kagome system.\cite{J.-C. DomengePRB2005,B. Fak 2012} Since the Kagome plane
of PbCu$_{3}$TeO$_{7}$ appear to have additional couplings ($2^{nd}$ and $%
3^{rd}$ $nn$), they might support ordered magnetic states as mentioned
above. In addition, there are also substantial inter-planar couplings found
in this system. These additional intra-Kagome and inter-Kagome couplings
might be the reason to have different AF transitions in this system. A
detailed neutron diffraction measurements and analysis will however be
needed to clarify the origin and nature of these transitions.

\section{Conclusions}

Based on the structural details, one forms the impression that PbCu$_{3}$TeO$%
_{7}$ might have quasi-$2D$\ staircase (buckled) Kagome layers. \ However,
our magnetic measurements on polycrystalline and single crystal samples show
the onset of LRO at about $36$ K with a change in the spin-order around $25$
K. This indicates the existence of significant three-dimensional couplings.
The $\chi (T)$ is found to obey the simple Curie-Weiss law above $200$ K
with a Curie constant as expected for a paramagnetic system. However, below $%
200$ K there is a deviation from the Curie-Weiss behavior though it is well
above $T_{N}$, which suggests the presence of frustrated spin correlations
well above $T_{N}$. This behavior is also consistent with $C_{p}(T)$ which
infers the presence of magnetic entropy well above the magnetic transitions.
We performed electronic structure calculations to determine the relative
values and importance of various exchange paths. Our results suggest the
presence of various intra-Kagome ($nnn$ and diagonal) and inter-Kagome
exchange hopping paths, which then must be responsible for the onset of LRO
in this system. Detailed neutron diffraction measurements will be helpful to
understand more about the nature of spin orderings and magnetic phase
diagram of this system.

\begin{acknowledgments}
The authors FCC and BKR acknowledge the support from National Science
Council of Taiwan under project number NSC-100-2119- M-002-021. JC thanks
CSIR India (Grant No. 09/080(0615)/2008-EMR -1) for research fellowship. AVM
and RK thank the Department of Science and Technology and CSIR, India for
financial support. Work at SNU was supported by National Creative Research
Initiative (2010-0018300).
\end{acknowledgments}

\textbf{Figures and captions:}

Fig. $1$ (a) The crystal structure of PbCu$_{3}$TeO$_{7}$ viewed along the $%
b-$direction is shown. (b) Projection of staircase (buckled) Kagome planes
in the $bc-$plane formed by Cu and O atoms with various bond-angles \cite{B
WedelZNB1996}. The possible nearest neighbor ($nn$) exchange couplings
between Cu1 and Cu2 atoms in the Kagome layer are denoted by $%
J_{1},J_{2},J_{3},$and $J_{4}$. The environments of the Cu1 octahedron, the
Cu2 tetrahedron and their coupling\ are shown in (c), (d), and (e). LDA
calculations suggest various hopping paths for (f)\ intra-Kagome plane and
(g) inter-Kagome planes indicated by $t_{n}$, and $t_{n}^{i},$ respectively.
(color online)

Fig. $\ 2$ (a) Powder x-ray diffraction pattern of PbCu$_{3}$TeO$_{7}$ at
room temperature. The open circles indicate the experimental data, while the
calculated XRD with residual factors of $R_{p}\approx 0.1,$ $R_{wp}\approx
0.12$, and $\chi ^{2}\approx 4$ is shown as a solid red line. The Bragg peak
positions are indicated by short-vertical marks and the bottom green line
represents the difference between the experimental and calculated data. (b)
The x-ray diffraction pattern when the x-rays are incident at the "top"
surface of the crystal in the horizontal plane (the incident and the
diffracted beam form the vertical plane). \ The crystallographic axes as
determined from Laue diffraction are also illustrated. (color online)

Fig. $3$ (a) Magnetic susceptibilities ($\chi )$ of a polycrystalline and a
single crystal sample are plotted as a function of $T.$ The inset shows an
expanded view of the low-temperature data. (b) ($\chi -\chi _{0})^{-1}$ is
plotted as a function of $T$. The data are fitted to the Curie-Weiss law in
the $T-$range $200$ K - $370$ K. (color online)

Fig. $4$ Magnetic susceptibility of PbCu$_{3}$TeO$_{7}$ in the orientation
of $H\parallel a$ for different magnetic fields (a) $1$ kOe, (b) $5$ kOe,
(c) $10$ kOe and (d) $15$ kOe is plotted. The black solid line indicates
zero-field-cooled (ZFC) warming data, while the red and blue solid lines are
the data related to field-cooled (FC) cooling and FC warming, respectively.
The red and black arrow marks indicate the position of magnetic anomaly in
the cooling and warming data, respectively. (color online)

Fig. $5$ The $C_{p}$ data (open circles) are plotted as a function of $T$.
The red solid line is a fit to the equation (see text) in the $T-$range from 
$150$ K to $280$ K and the red dashed line is the extrapolation of the fit.
Inset shows magnetic heat capacity (left $y-$axis) and entropy (right $y-$%
axis) versus $T$. (color online)

Fig. $6$ Downfolded band structure (shown in red line) compared to the full
orbital band structure (shown in black line) for PbCu$_{3}$TeO$_{7}$. (color
online)

Fig. $7$ Partial density of states obtained from LMTO-ASA. (color online)

Fig. $8$ Wannier function of Cu-d$_{x^{2}-y^{2}}$ placed at Cu2 site. (color
online)

\bigskip

\textbf{Tables and captions:}

%TCIMACRO{\TeXButton{B}{\begin{table}[h]\centering}}%
%BeginExpansion
\begin{table}[h]\centering%
%EndExpansion
%TCIMACRO{%
%\TeXButton{Caption}{\caption{Atomic coordinates, occupancy factors for PbCu$_{3}$TeO$_{7}$}}}%
%BeginExpansion
\caption{Atomic coordinates, occupancy factors for PbCu$_{3}$TeO$_{7}$}%
%EndExpansion
\begin{tabular}{|l|l|l|l|l|l|}
\hline
Atom & Wyckoff site & x & y & z & occupancy \\ \hline
Pb & 4c & 0.10691 & 0.25 & 0.54683 & 1 \\ \hline
Te & 4c & 0.38594 & 0.25 & 0.30011 & 1 \\ \hline
Cu1 & 8d & 0.38687 & 0.01543 & 0.62710 & 1 \\ \hline
Cu2 & 4c & 0.24769 & 0.25 & 0.89997 & 1 \\ \hline
O1 & 4c & 0.21017 & 0.25 & 0.15840 & 1 \\ \hline
O2 & 4c & 0.30574 & 0.25 & 0.68569 & 1 \\ \hline
O3 & 4c & 0.51777 & 0.25 & 0.4605 & 1 \\ \hline
O4 & 8d & 0.18466 & -0.05591 & 0.99033 & 1 \\ \hline
O5 & 8d & 0.4347 & -0.06281 & 0.64738 & 1 \\ \hline
\end{tabular}%
%TCIMACRO{\TeXButton{E}{\end{table}}}%
%BeginExpansion
\end{table}%
%EndExpansion

%TCIMACRO{\TeXButton{B}{\begin{table}[h]\centering}}%
%BeginExpansion
\begin{table}[h]\centering%
%EndExpansion
%TCIMACRO{%
%\TeXButton{Caption}{\caption{The  parameters obtained from the Curie-Weiss fit in the temperature range from 200 K to 370 K}}}%
%BeginExpansion
\caption{The  parameters obtained from the Curie-Weiss fit in the temperature range from 200 K to 370 K}%
%EndExpansion
\begin{tabular}{|c|c|c|l|l|}
\hline
& $T-$range (K) & $C$ (cm$^{3}$ K/mol Cu) & $\theta _{CW}$ (K) & $\mu _{eff}$
($\mu _{B})$ \\ \hline
$H\parallel a$ & $200$ - $370$ & $0.52\pm 0.05$ & $-(145\pm 5)$ & $2.04$ \\ 
\hline
$H\parallel b$ & $200$ - $370$ & $0.49\pm 0.05$ & $-(155\pm 5)$ & $1.98$ \\ 
\hline
$H\parallel c$ & $200$ - $370$ & $0.46\pm 0.05$ & $-(140\pm 5)$ & $1.92$ \\ 
\hline
\end{tabular}%
%TCIMACRO{\TeXButton{E}{\end{table}}}%
%BeginExpansion
\end{table}%
%EndExpansion

%TCIMACRO{\TeXButton{B}{\begin{table}[h]\centering}}%
%BeginExpansion
\begin{table}[h]\centering%
%EndExpansion
%TCIMACRO{%
%\TeXButton{Caption}{\caption{The variation of magnetic peak positions of T$_{N3}$ in the cooling and warming $M(T)$  data in different magnetic fields}}}%
%BeginExpansion
\caption{The variation of magnetic peak positions of T$_{N3}$ in the cooling and warming $M(T)$  data in different magnetic fields}%
%EndExpansion
\begin{tabular}{|c|c|c|c|}
\hline
$H$ (kOe) & $T_{N3}$ (K) (FC cooling) & $T_{N3}$ (K) (ZFC\ warming) & 
Difference $\Delta T$ (K) \\ \hline
1 & $15.8\pm 0.1$ & $16.5\pm 0.1$ & $0.7$ K \\ \hline
5 & $16.3\pm 0.1$ & $16.8\pm 0.1$ & $0.5$ K \\ \hline
10 & $17.2\pm 0.1$ & $17.5\pm 0.1$ & $0.3$ K \\ \hline
15 & $17.5\pm 0.1$ & $17.7\pm 0.1$ & $0.2$ K \\ \hline
\end{tabular}%
%TCIMACRO{\TeXButton{E}{\end{table}}}%
%BeginExpansion
\end{table}%
%EndExpansion

%TCIMACRO{\TeXButton{B}{\begin{table}[h]\centering}}%
%BeginExpansion
\begin{table}[h]\centering%
%EndExpansion
%TCIMACRO{%
%\TeXButton{Caption}{\caption{Hopping (in meV) between various Cu atoms within the Kagome plane, obtained from NMTO downfolding method.}}}%
%BeginExpansion
\caption{Hopping (in meV) between various Cu atoms within the Kagome plane, obtained from NMTO downfolding method.}%
%EndExpansion
\begin{tabular}{|c|c|c|c|p{4cm}|}
\hline\hline
Coupling between & Hopping path & Distance (\AA ) & Hopping (meV) & 
Bond-angle and bond-lengths \\ \hline\hline
Cu1 - Cu2 $(nn)$ & $t_{1}$ & 2.90 & 46.23 & Cu1-O4-Cu2$\sim $ 92.9$%
%TCIMACRO{\U{b0}}%
%BeginExpansion
{{}^\circ}%
%EndExpansion
$, Cu1-O1-Cu2$\sim $ 83.1$%
%TCIMACRO{\U{b0}}%
%BeginExpansion
{{}^\circ}%
%EndExpansion
$ \\ \hline
Cu1 - Cu1 $(nn)$ & $t_{2}$ & 3.065 & 42.17 & Cu1-O2-Cu1$\sim $105$%
%TCIMACRO{\U{b0}}%
%BeginExpansion
{{}^\circ}%
%EndExpansion
$ \\ \hline
Cu1 - Cu2 $(nn)$ & $t_{3}$ & 3.264 & 149.67 & Cu1-O2-Cu2$\sim $118.1$%
%TCIMACRO{\U{b0}}%
%BeginExpansion
{{}^\circ}%
%EndExpansion
$ \\ \hline
Cu1 - Cu1 $(nn)$ & $t_{4}$ & 3.29 & 91.16 & Cu1-O3-Cu1$\sim $106.7$%
%TCIMACRO{\U{b0}}%
%BeginExpansion
{{}^\circ}%
%EndExpansion
,$ Cu1-O1-Cu1$\sim $91$%
%TCIMACRO{\U{b0}}%
%BeginExpansion
{{}^\circ}%
%EndExpansion
$ \\ \hline
Cu2 - Cu2 $(nnn)$ & $t_{5}$ & 5.43 & 57 & O1-O4$\sim $2.8 \AA , Cu2-O1-O4$%
\sim $ 152$%
%TCIMACRO{\U{b0}}%
%BeginExpansion
{{}^\circ}%
%EndExpansion
,$O1-O4-Cu2-$\sim $ 110$%
%TCIMACRO{\U{b0}}%
%BeginExpansion
{{}^\circ}%
%EndExpansion
$ \\ \hline
Cu1 - Cu2 $\ (nnn)$ & $t_{6}$ & 5.617 & 19.05 & O1-O5$\sim $2.78 \AA  \\ 
\hline
Cu2 - Cu2 (diagonal) & $t_{7}$ & 6.353 & 104.76 & O4-O4$\sim $2.57 \AA $,$%
Cu2-O4-O4$\sim $ 167$%
%TCIMACRO{\U{b0}}%
%BeginExpansion
{{}^\circ}%
%EndExpansion
,$O4-O4-Cu2$\sim $ 167$%
%TCIMACRO{\U{b0}}%
%BeginExpansion
{{}^\circ}%
%EndExpansion
$ \\ \hline
Cu1 - Cu2 (long-range) & $t_{8}$ & 6.7 & 49 & O-O$\sim $2.83 \AA  \\ \hline
\end{tabular}%
%TCIMACRO{\TeXButton{E}{\end{table}}}%
%BeginExpansion
\end{table}%
%EndExpansion

%TCIMACRO{\TeXButton{B}{\begin{table}[h]\centering}}%
%BeginExpansion
\begin{table}[h]\centering%
%EndExpansion
%TCIMACRO{%
%\TeXButton{Caption}{\caption{Hopping parameters (in meV) for inter Kagome plane, obtained from NMTO downfolding method.}}}%
%BeginExpansion
\caption{Hopping parameters (in meV) for inter Kagome plane, obtained from NMTO downfolding method.}%
%EndExpansion
\begin{tabular}{|c|c|c|c|c|}
\hline\hline
\multicolumn{1}{|c|}{Coupling between} & \multicolumn{1}{|c|}{Hopping path}
& \multicolumn{1}{|c|}{Distance (\AA )} & \multicolumn{1}{|c|}{Hopping (meV)}
& Bond-angle and bond-lengths \\ \hline\hline
Cu2 - Cu2 & $t_{1}^{i}$ & 5.92 & 16.32 &  \\ \hline
Cu1 - Cu2 & $t_{2}^{i}$ & 6.12 & 49 & Cu1-O3-Te-O4-Cu2, O3-O4$\sim $2.72 \AA 
\\ \hline
Cu1 - Cu2 & $t_{3}^{i}$ & 6.27 & 20.41 &  \\ \hline
Cu2 - Cu2 & $t_{4}^{i}$ & 6.27 & 28.53 &  \\ \hline
\end{tabular}%
%TCIMACRO{\TeXButton{E}{\end{table}}}%
%BeginExpansion
\end{table}%
%EndExpansion

\bigskip

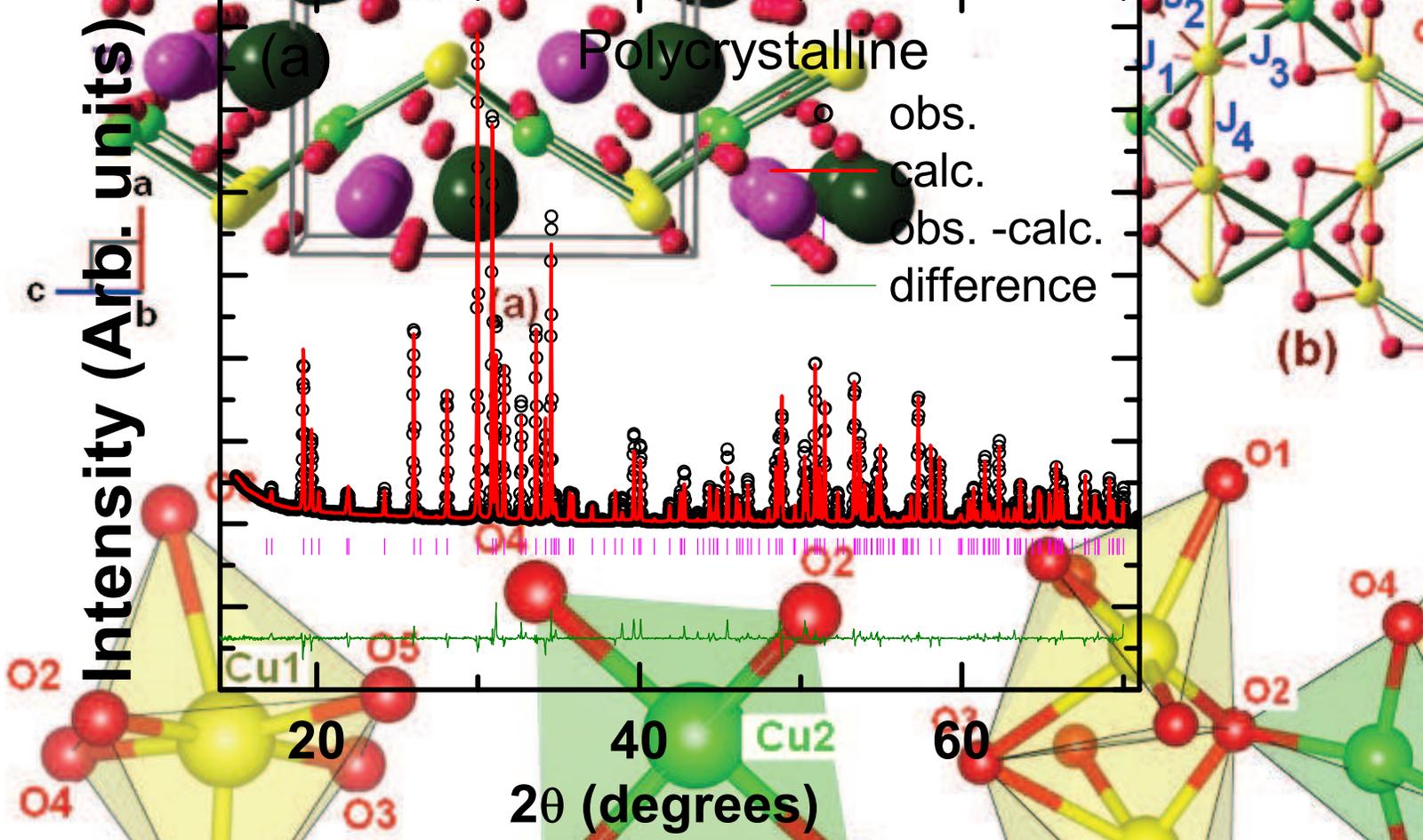
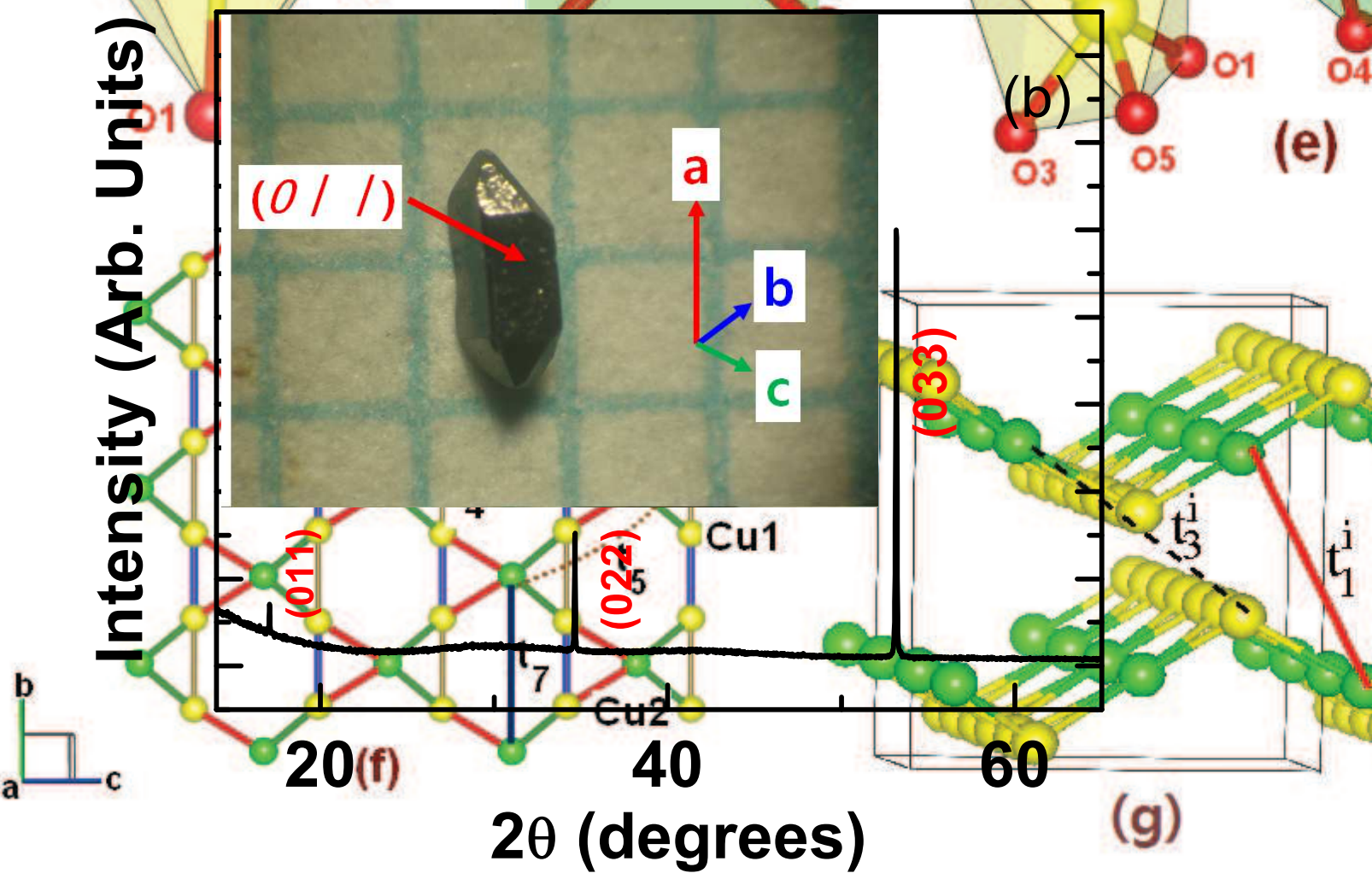
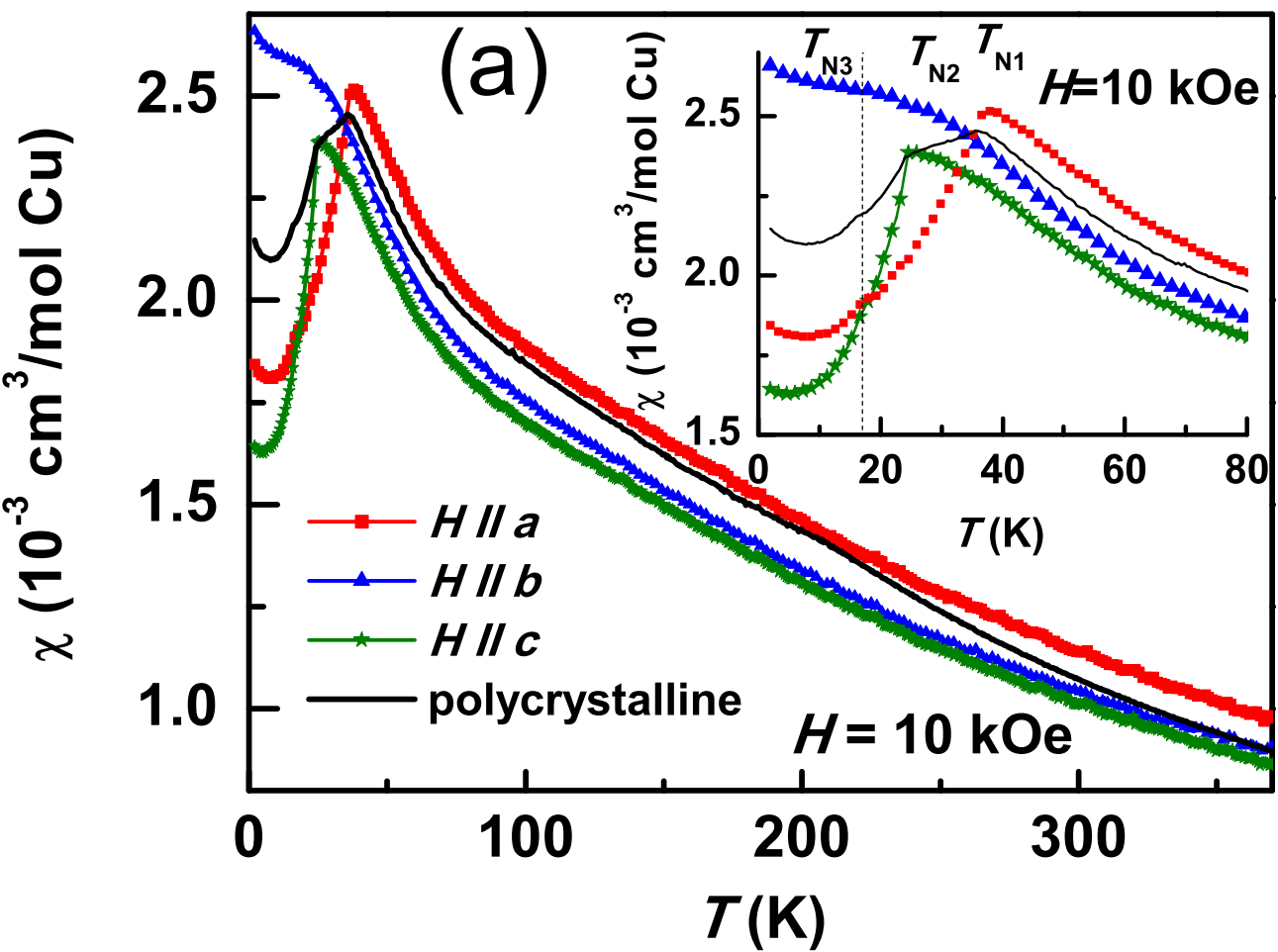
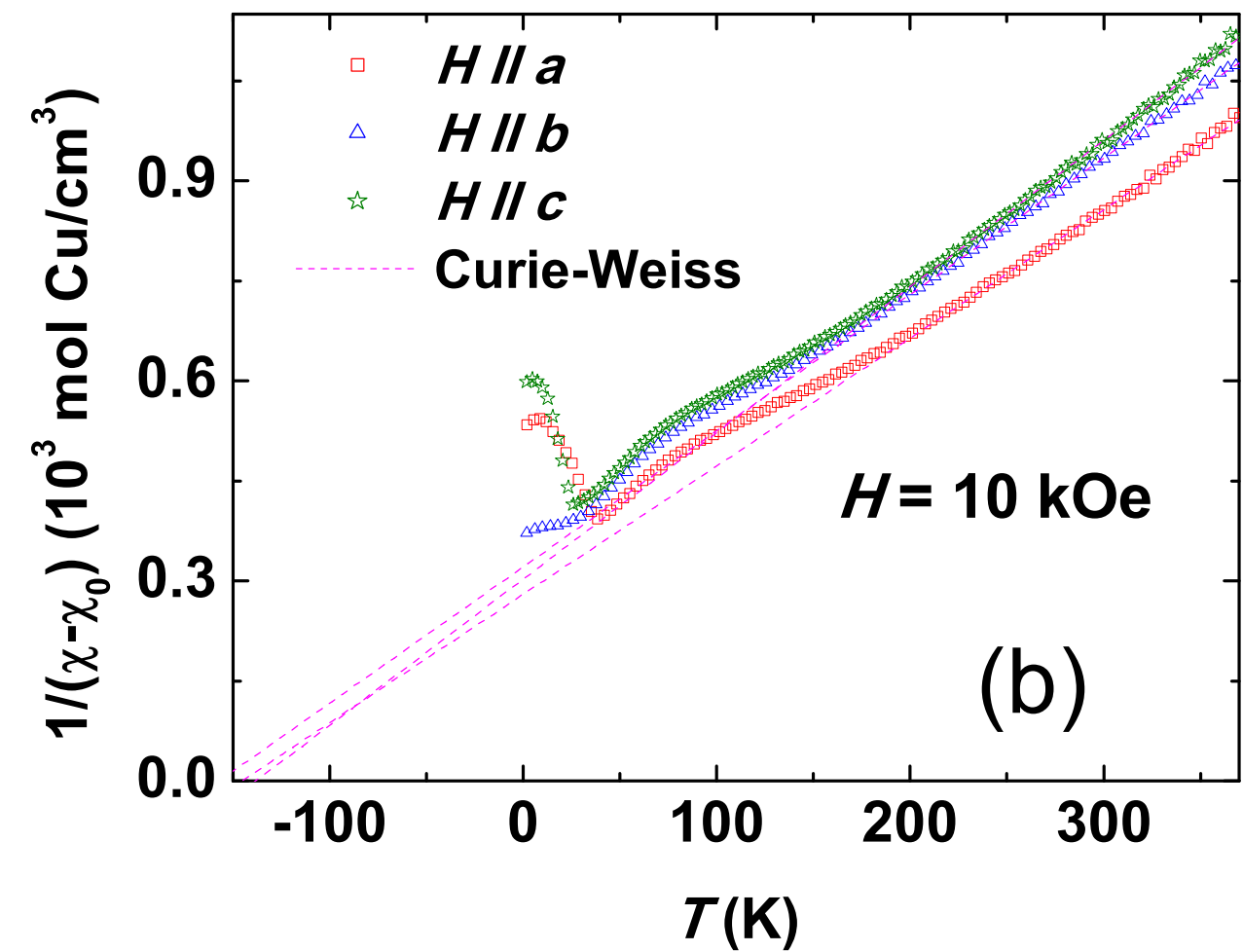
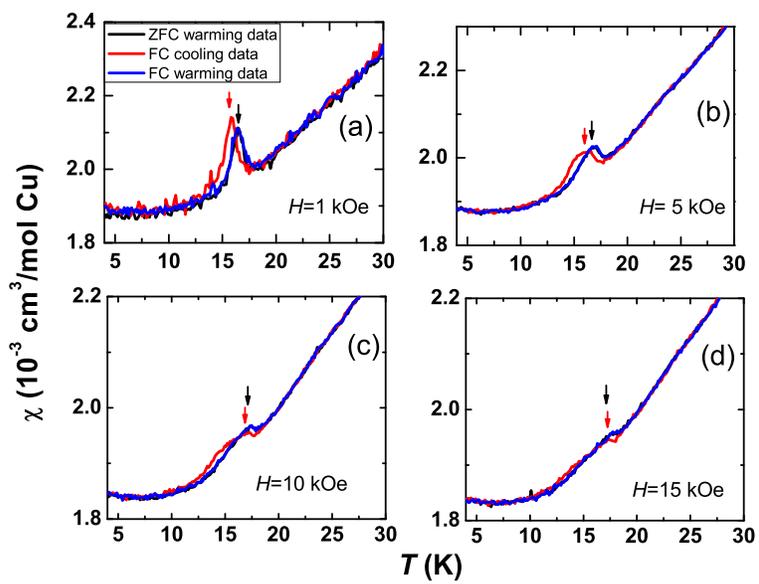
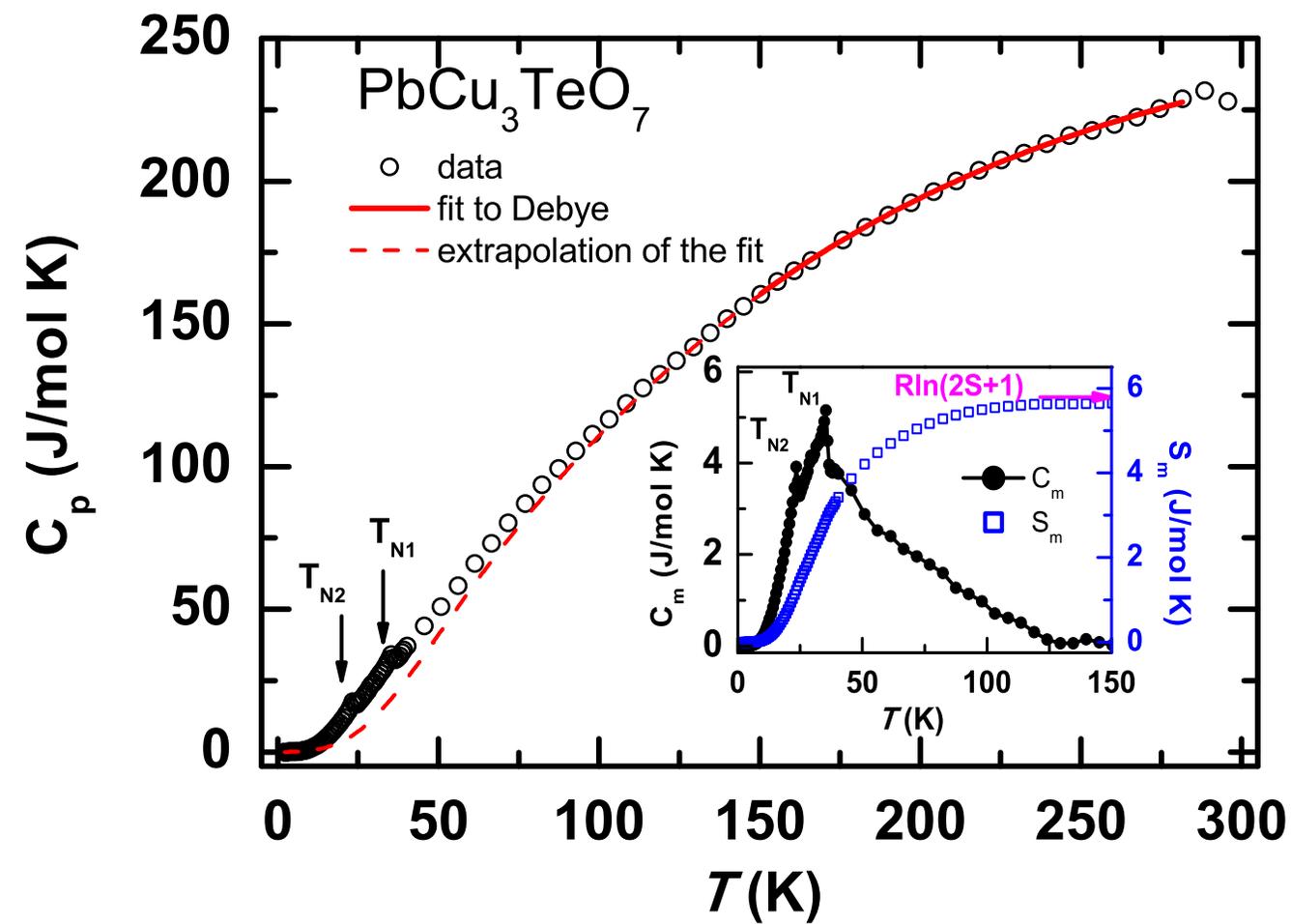
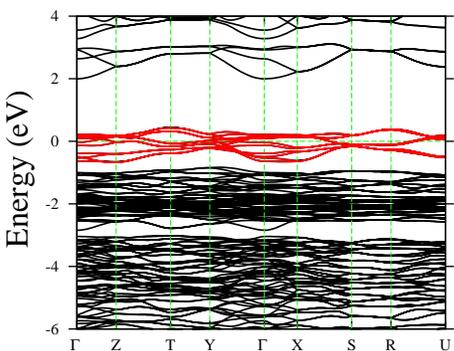
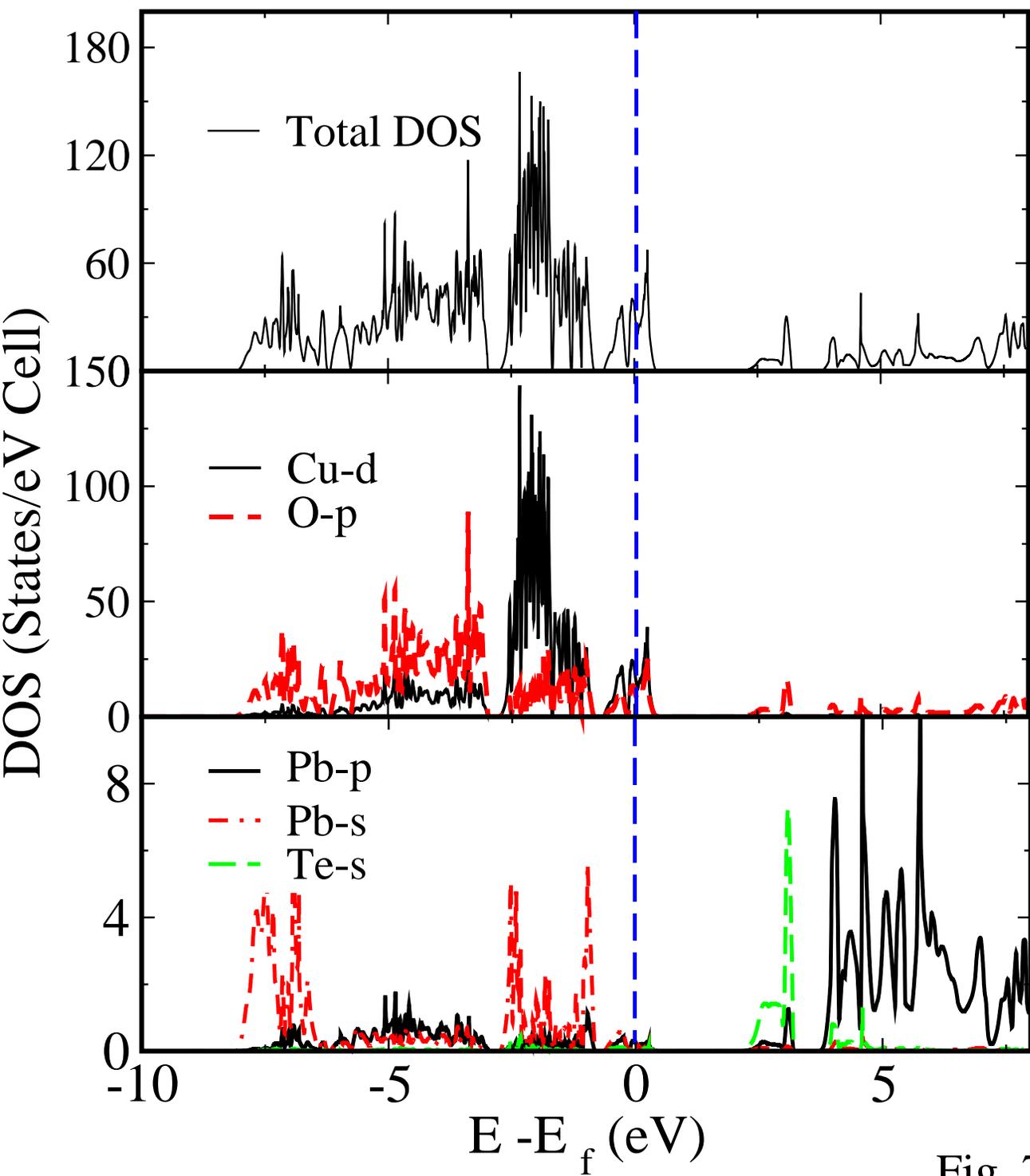
\end{document}